\documentclass[conference, letterpaper]{IEEEtran}
\IEEEoverridecommandlockouts
\usepackage{amssymb,amsfonts}
\usepackage[cmex10]{amsmath}
\usepackage[utf8]{inputenc}
\usepackage[T1]{fontenc}
\usepackage{url}

\usepackage{algorithmic}
\usepackage{bbm}
\usepackage{bm}
\usepackage{cite}
\usepackage{graphicx}
\usepackage{textcomp}
\usepackage{xcolor}

\usepackage{hyperref}
\hypersetup{hidelinks, 
colorlinks=true,
allcolors=black,
pdfstartview=Fit,
breaklinks=true}

\def\BibTeX{{\rm B\kern-.05em{\sc i\kern-.025em b}\kern-.08em
    T\kern-.1667em\lower.7ex\hbox{E}\kern-.125emX}}
\begin{document}

\bibliographystyle{IEEEtran}
\title{On Finite-Time Mutual Information}
    \author{\IEEEauthorblockN{Jieao~Zhu, Zijian~Zhang, Zhongzhichao~Wan, and~Linglong~Dai}
    \IEEEauthorblockA{
        Beijing National Research Center for Information Science and Technology (BNRist) \\
        Department of Electronic Engineering, Tsinghua University, Beijing 100084, China \\
        E-mails: zja21, zhangzj20, wzzc20@mails.tsinghua.edu.cn; daill@tsinghua.edu.cn
    }
}

\maketitle

\begin{abstract}
    Shannon-Hartley theorem can accurately calculate the channel capacity when the signal observation time is infinite. However, the calculation of finite-time mutual information, which remains unknown, is essential for guiding the design of practical communication systems. In this paper, we investigate the mutual information between two correlated Gaussian processes within a finite-time observation window. We first derive the finite-time mutual information by providing a limit expression. Then we numerically compute the mutual information within a single finite-time window. We reveal that the number of bits transmitted per second within the finite-time window can exceed the mutual information averaged over the entire time axis, which is called the exceed-average phenomenon. Furthermore, we derive a finite-time mutual information formula under a typical signal autocorrelation case by utilizing the Mercer expansion of trace class operators, and reveal the connection between the finite-time mutual information problem and the operator theory. Finally, we analytically prove the existence of the exceed-average phenomenon in this typical case, and demonstrate its compatibility with the Shannon capacity.
\end{abstract}


% Setup Theorem/Lemma/Remark environments.
\newtheorem{theorem}{\bf Theorem}
\newtheorem{lemma}{\bf Lemma}
\newtheorem{remark}{Remark}

\section{Introduction}
The Shannon-Hartley theorem \cite{Shannon} has accurately revealed the fundamental theoretical limit of information transmission rate $C$, which is also called as the Shannon capacity, over a Gaussian waveform channel of a limited bandwidth $W$. The expression for Shannon capacity is $C=W\log\left(1+S/N\right)$, where $S$ and $N$ denote the signal power and the noise power, respectively. The derivation of Shannon-Hartley Theorem heavily depends on the Nyquist sampling principle \cite{Sampling1967Landau}. The Nyquist sampling principle, which is also named as the $2WT$ theorem \cite{Nyquist1928CertainTopics}, claims that one can only obtain $2WT + o(2WT)$ independent samples within an observation time window $T$ in a channel band-limited to $W$ \cite{BW1976Slepian}.

Based on the Nyquist sampling principle, the Shannon capacity is derived by multiplying the capacity $1/2 \log (1+P / N)$ of a Gaussian symbol channel \cite[p.249]{Cover1999ElementsInfTheory} with $2WT+o(2WT)$ at first, and then dividing the result by $T$, finally letting $T\rightarrow \infty$. Note that this approximation only holds when $T \to \infty$. Therefore, the Shannon capacity only asymptotically holds as $T$ becomes sufficiently large. When $T$ is of finite value, the approximation fails to work. Thus, when the observation time $T$ is finite, the Shannon-Hartley Theorem cannot be directly applied to calculate the capacity in a finite-time window. To the best of our knowledge, the evaluation of the finite-time mutual information has not yet been investigated in the literature. It is worth noting that real-world communication systems transmit signals in a finite-time window, thus evaluating the finite-time mutual information is of practical significance. \par 

% Our contributions
In this paper, to fill in this gap, we analyze the finite-time mutual information instead of the traditional infinite-time counterpart, and prove the existence of exceed-average phenomenon within a finite-time observation window\footnote{Simulation codes will be provided to reproduce the results presented in this paper: \href{http://oa.ee.tsinghua.edu.cn/dailinglong/publications/publications.html}{http://oa.ee.tsinghua.edu.cn/dailinglong/publications/publications.html}.}. Specifically, our contributions are summarized as follows:

% Contributions: concise, but full of insights.
\begin{itemize}
	\item We derive the mutual information expressions within a finite-time observation window by using dense sampling and limiting methods. In this way, we can overcome the continuous difficulties that appear when analyzing the information contained in a continuous time interval. These finite-time mutual information expressions make the analysis of finite-time  problems possible.
    \item We conduct numerical experiments based on the discretized formulas. In the numerical results under a special setting, we reveal the exceed-average phenomenon, i.e., the mutual information within a finite-time observation window exceeds the mutual information averaged over the whole time axis.
    \item In order to analytically prove the exceed-average phenomenon, we first derive an analytical finite-time mutual information formula based on Mercer expansion \cite{mercer1909functions}, where we can find the connection between the mutual information problem and the operator theory \cite{zhu2007operator}. To make the problem tractable, we construct a typical case in which the transmitted signal has certain statistical properties. Utilizing this construction, we obtain a closed-form mutual information solution in this typical case, which leads to a rigorous proof of the exceed-average phenomenon.
    % Add insight.
\end{itemize}

% Paper organization. 
{\it Organization}: In the rest of this paper, the finite-time mutual information is formulated and evaluated numerically in Section II, where the exceed-average phenomenon is first discovered. Then, in Section III, we derive a closed-form finite-time mutual information formula under a typical case. Based on this formula, in Section IV, the exceed-average phenomenon is rigorously proved. Finally, conclusions are drawn in Section V.

% Symbol Table.
{\it Notations}: $X(t)$ denotes a Gaussian Process; $R_X(t_1, t_2)$ denotes the autocorrelation function; $S_X(f), S_X(\omega)$ are the Power Spectral Density (PSD) of the corresponding process $X(t)$; Boldface italic symbols $\bm X(t_1^n)$ denotes the column vector generated by taking samples of  $X(t)$ on instants $t_i, 1 \leq i \leq n$; Upper-case boldface letters such as ${\bf{\Phi}}$ denote matrices; $\mathbb{E\left[\cdot\right]}$ denotes the expectation; $\mathbbm{1}_{A}(\cdot)$ denotes the indicator function of the set $A$; $\mathcal{L}^2([0,T])$ denotes the collection of all the square-integrable functions on window $[0,T]$; ${\rm i}$ denotes the imaginary unit.

\section{Numerical Analysis of the Finite-Time Mutual Information}
	In this section, we numerically evaluate the finite-time mutual information. In Subsection~\ref{sect_2_subsect_1}, we model the transmission problem by Gaussian processes, and derive the mutual information expressions within a finite-time observation window; In Subsection~\ref{sect_2_subsect_2}, we approximate the finite-time mutual information by discretized matrix-based formulas; In Subsection~\ref{sect_2_subsect_3}, we reveal the exceed-average phenomenon by numerically evaluating the finite-time mutual information.

	\subsection{The Expressions for finite-time mutual information}\label{sect_2_subsect_1}
    Inspired by \cite{GPSampling1957BA}, we model the transmitted signal by a zero-mean stationary Gaussian stochastic process, denoted as $X(t)$, and the received signal by $Y(t):=X(t)+N(t)$. The noise process $N(t)$ is also a stationary Gaussian process independent of $X(t)$. The receiver can only observe the signal within a finite-time window $[0,T]$, where $T>0$ is the observation window span. We aim to find the maximum mutual information that can be acquired within this time window. 

    To analytically express the amount of acquired information, we first introduce $n$ sampling points inside the time window, denoted by $(t_1, t_2 , \cdots, t_n):=t_1^n$, and then let $n \rightarrow \infty$ to approximate the finite-time mutual information. By defining ${\bm X}(t_1^n)\equiv (X(t_1), X(t_2), \cdots, X(t_n))$ and ${\bm Y}(t_1^n)\equiv (Y(t_1), Y(t_2), \cdots, Y(t_n))$, the mutual information on these $n$ samples can be expressed as
    \begin{equation}
        I(t_1^n) = I({\bm X}(t_1^n); {\bm Y}(t_1^n)),
        \label{eq_sample_capacity}
    \end{equation}
    and the finite-time mutual information is defined as
    \begin{equation}
        I(T) = \lim_{n\rightarrow \infty}\sup_{\{t_1^n\} \subset [0,T]}{I(t_1^n)}.
        \label{eq_finite_time_capacity}
    \end{equation}
    Then, the transmission rate $C(T)$ can be defined as $C(T)=I(T)/T$. After these definitions, we can then define the limit mutual information as $C(\infty)=\lim_{T\rightarrow \infty}{C(T)}$ by letting $T\rightarrow \infty$.

    \subsection{Discretization}\label{sect_2_subsect_2}
    Without loss of generality, we fix the sampling instants uniformly onto fractions of $T$: $t_i = (i-1)T/n, 1\leq i \leq n$. Since the random vectors ${\bm X}(t_1^n)$ and ${\bm Y}(t_1^n)$ are samples of a Gaussian process, they are both Gaussian random vectors with mean zero and covariance matrices ${\bm K}_X$ and ${\bm K}_Y$, where ${\bm K}_X, {\bm K}_Y \in \mathbb{R}^{n\times n}$ are symmetric positive-definite matrices defined as
    \begin{equation}
        \begin{aligned}
            ({\bm K}_X)_{i,j} &= R_X(t_i, t_j) := \mathbb{E}\left[X(t_i)X(t_j)\right], \\
            ({\bm K}_Y)_{i,j} &= R_Y(t_i, t_j) := \mathbb{E}\left[Y(t_i)Y(t_j)\right]. \\
        \end{aligned}
    \end{equation}
    Note that $Y(t)$ is the independent sum of $X(t)$ and $N(t)$, thus the autocorrelation functions satisfy $R_Y(t_1,t_2)=R_X(t_1,t_2)+R_N(t_1,t_2)$, and similarly the covariance matrices satisfy ${\bm K}_Y={\bm K}_X+{\bm K}_N$.
    
    The mutual information $I(t_1^n)$ is defined as $I(t_1^n)=h({\bm Y}(t_1^n))-h({\bm Y}(t_1^n)|{\bm X}(t_1^n))=h({\bm Y}(t_1^n))-h({\bm N}(t_1^n))$, where $h(\cdot)$ denotes the differential entropy. Utilizing the entropy formula for $n$-dimentional Gaussian vectors \cite{Cover1999ElementsInfTheory}, we obtain 
    \begin{equation}
        I(t_1^n) = \frac{1}{2} \log\left(\frac{\det({\bm K}_X+{\bm K}_N)}{\det{({\bm K}_N)}} \right).
        \label{eq_I_expression}
    \end{equation}

    \subsection{Numerical Analysis}\label{sect_2_subsect_3}
        In order to study the properties of mutual information $I(t_1^n)$ as a function of $n$, we set the autocorrelation function of the signal process $X(t)$ and noise process $N(t)$ to the following special case
        \begin{equation}
                \begin{aligned}
                    R_X(t_1, t_2) & = \mathrm{sinc}(10(t_1-t_2)),            \\
                    R_N(t_1, t_2) & = \exp(-\lvert t_1-t_2\rvert),           \\
                \end{aligned}
            \label{eq_correlation_examples_1}
        \end{equation}
        where $\mathrm{sinc}(x):=\sin(\pi x)/(\pi x)$, and the corresponding PSDs are $S_X(f)  = 0.1\times \mathbbm{1}_{\{-5\leq f \leq 5\}}$ and $S_N(f) = \frac{2}{1+(2\pi f)^2}$. In order to compare the finite-time mutual information with the averaged mutual information, the Shannon limit with colored noise spectrum $S_N(f)$ is utilized, which is a generalized version of the well-known formula $C=W\log\left(1+S/N\right)$. The averaged mutual information of colored noise PSD \cite{Cover1999ElementsInfTheory} is expressed as 
        \begin{equation}
            C_{\rm av} := \frac{1}{2} 
            \int_{-\infty}^{+\infty}{\log\left(1+\frac{S_X(f)}{S_N(f)}\right)\mathrm{d}
             f}.
            \label{eq_shannon_formula}
        \end{equation}
        Then, plugging \eqref{eq_correlation_examples_1} into \eqref{eq_shannon_formula} yields the numerical result for $C_{\rm av}$.
        
\begin{figure}[!t]
            \centering
            \includegraphics[width=8.5cm]{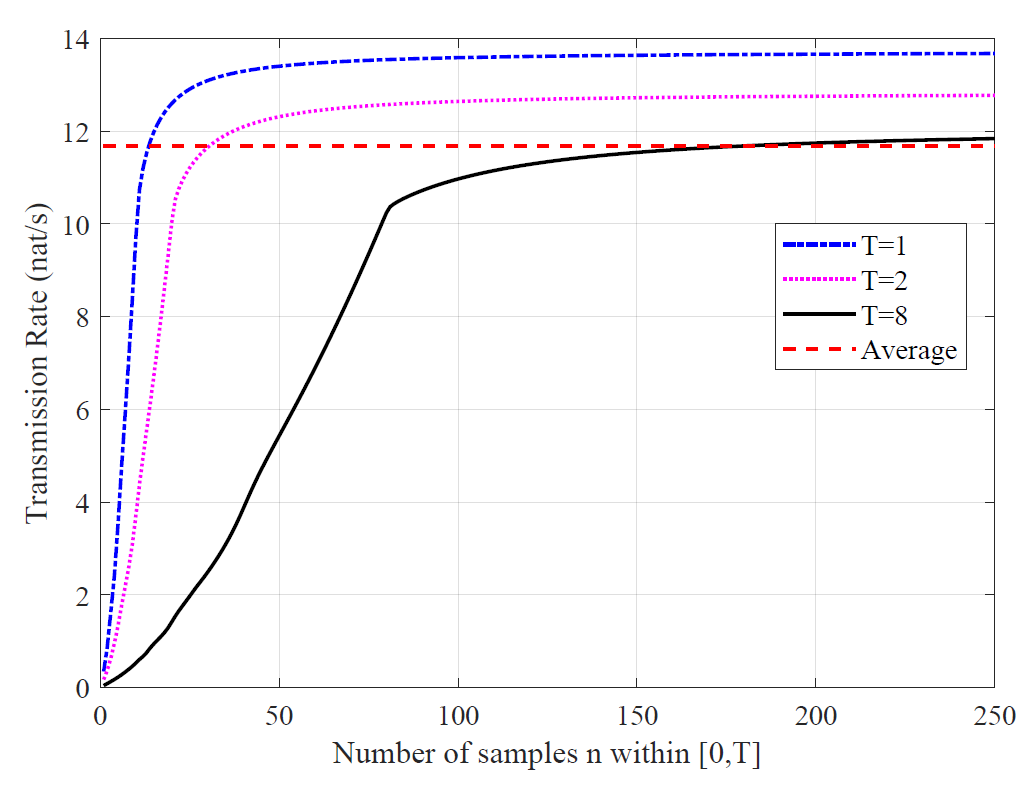}
            \caption{A first glance to the exceed-average phenomenon. The red dashed horizontal line is the Shannon limit, and the $T=1, 2, 8$ curves illustrate the dependence of $C(t_1^n)$ on $n$.}
            \label{fig_sp_finite_time_rate_wrt_n}
        \end{figure}

        We calculate the finite-time transmission rate $C(T)$ and the average mutual information $C_{\rm av}$ against the number of samples $n$ within the observation window $[0,T]$. The numerical results are collected in Fig.~\ref{fig_sp_finite_time_rate_wrt_n}. It is shown that $I(t_1^n)$ is an increasing function of $n$, and for fixed values of $T$, the approximated finite-time mutual information $I(t_1^n)$ tends to a finite limit under the correlation assumptions given by \eqref{eq_correlation_examples_1}. The most amazing observation is that, it is possible to obtain more information within finite-time window $[0,T]$ than the prediction $T C_{\rm av}$ given by averaging the mutual information along the entire time axis \eqref{eq_shannon_formula}. We call this phenomenon the exceed-average phenomenon.

\section{A Closed-Form Finite-Time Mutual Information Formula}
    In this section, we first introduce the Mercer expansion in Subsection~\ref{sect_3_subsect_1} as a basic tool for our analysis. Then we derive the series representation of the finite-time mutual information, and the corresponding power constraint in Subsection~\ref{sect_3_subsect_2}, under the assumption of AWGN noise.
	
    \subsection{The Mercer Expansion}\label{sect_3_subsect_1}
    By analyzing the structure of the autocorrelation functions, it is possible to obtain $I(t_1^n)$ and $I(T)$ directly. In fact, if we know the Mercer expansion \cite{mercer1909functions} of the autocorrelation function $R_X(t_1, t_2)$ on interval $[0,T]$, then we can calculate $h({\bm X}(t_1^n))$ more easily \cite{MercerExpansion1955Barrett}. In the following discussion, we assume the Mercer expansion of the source autocorrelation function $R_X(t_1, t_2)$ to be in the following form
    \begin{equation}
        \lambda_k \phi_k(t_1) = \int_{0}^{T}{R_X(t_1, t_2) \phi_k(t_2) \mathrm{d}t_2}; k > 0, k\in \mathbb{N},
        \label{eq_mercer_expansion}
    \end{equation}
    where the eigenvalues are positive: $\lambda_k > 0$, and the eigenfunctions form an orthonormal set:
    \begin{equation}
        \int_{0}^{T}{\phi_i(t)\phi_j(t)\mathrm{d}t} = \delta_{ij}.
        \label{eq_orthonormal}
    \end{equation}
        
    The Mercer's theorem \cite{mercer1909functions} ensures the existence and uniqueness of the eigenpairs $(\lambda_k, \phi_k(t))_{k=1}^{\infty}$, and furthermore, the kernel itself can be expanded under the eigenfunctions:
    \begin{equation}
        R_X(t_1, t_2) = \sum_{k=1}^{+\infty}{\lambda_k \phi_k(t_1)\phi_k(t_2)}.
        \label{eq_mercer_kernel_expansion}
    \end{equation}
    
    \subsection{Finite-Time Mutual Information Formula} \label{sect_3_subsect_2}
    Based on Mercer expansion, we obtain a closed-form formula in the following {\bf Theorem \ref{th_1}}.
        \begin{theorem}[Source expansion, AWGN noise]\label{th_1}
            Suppose the information source, modeled by the transmitted process $X(t)$, has autocorrelation function $R_X(t_1, t_2)$. An AWGN noise of PSD $n_0/2$ is imposed onto $X(t)$, resulting in the received process $Y(t)$. The Mercer expansion of $R_X(t_1, t_2)$ on $[0,T]$ is given by \eqref{eq_mercer_expansion}, satisfying \eqref{eq_orthonormal}. Then the finite-time mutual information $I(T)$ within the observation window $[0,T]$ between the processes $X(t)$ and $Y(t)$ can be expressed as
            \begin{equation}
                I(T)=\frac{1}{2} \sum_{k=1}^{+\infty}{\log\left(1+\frac{\lambda_k}{n_0/2}\right)}.
                \label{eq_th_1}
            \end{equation}
        \end{theorem}
    	\begin{IEEEproof}
    		The Mercer expansion decomposes the AWGN channel into an infinite number of independent parallel subchannels, each with signal power $\lambda_k$ and noise variance $n_0/2$. Thus, accumulating all the mutual information values of these subchannels yields the finite-time mutual information $I(T)$. 
    	\end{IEEEproof}
        From {\bf Theorem \ref{th_1}} we can conclude that the finite-time mutual information of AWGN channel is uniquely determined by the Mercer spectra $\lambda_k$ of $R_X(t_1, t_2)$ within $[0,T]$. However, it remains unknown whether the series representation \eqref{eq_th_1} converges. In fact, the convergence is closely related to the signal power, which is calculated in the following {\bf Lemma \ref{lemma_1}}.
            \begin{lemma}[Operator Trace Coincide with Power Constraint]
                Given stationary Gaussian process $X(t)$ with mean zero and autocorrelation $R_X(t_1, t_2)$. The Mercer expansion of $R_X(t_1, t_2)$ on $[0,T]$ is given by \eqref{eq_mercer_expansion}, satisfying \eqref{eq_orthonormal}. The Mercer operator $M(\cdot): \mathcal{L}^2([0,T]) \rightarrow \mathcal{L}^2([0,T])$ is defined by the integral $(M\phi)(s) = \int_{0}^{T}{R_X(s ,\tau)\phi(\tau)\mathrm{d}\tau}$. Then the sum of all the eigenvalues $\lambda_k$ of operator $M$ is equal to the signal energy $PT$ within $[0,T]$:
                \begin{equation}
                    \mathrm{tr}(M) := \sum_{k=1}^{+\infty}{\lambda_k} = PT,
                \end{equation}
                where $P=R_X(0,0)$.
                \label{lemma_1}
            \end{lemma}
        	\begin{IEEEproof}
        		Integrating both sides of \eqref{eq_mercer_kernel_expansion} on the interval $[0,T]$ gives the conclusion immediately.
        	\end{IEEEproof}
			\begin{remark}\label{remark_1}
				From the above {\bf Lemma \ref{lemma_1}}, we can conclude that the sum of $\lambda_k$ is finite when $T$ is finite. It can be immediately derived that $I(T)<\infty$, since $\log(1+x) \leq x$. 
            \end{remark}
            \begin{remark}
                The finite-time mutual information formula~\eqref{eq_th_1} is closely related to the operator theory \cite{zhu2007operator} in functional analysis. The sum of all the eigenvalues $\lambda_k$ is called the operator trace in linear operator theory. As is mentioned in {\bf Lemma \ref{lemma_1}}, the autocorrelation function $R_X(t_1, t_2)$ can be treated as a linear operator $M$ on $\mathcal{L}^2([0,T])$.
			\end{remark}

\section{Proof of the Existence of Exceed-Average Phenomenon}
    In this section, we first give a proof of the existence of the exceed-average phenomenon in a typical case, then we discuss the compatibility with the Shannon-Hartley theorem.

    \subsection{Closed-Form Mutual Information in a Typical Case} \label{sect_4_subsect_1}
        To prove the exceed-average phenomenon, we only need to show that the finite-time mutual information is greater than the averaged mutual information in a typical case. Let us consider a finite-time communication scheme with a finitely-powered stationary transmitted signal autocorrelation\footnote{The signal autocorrelation $R_X(\tau)$ is often observed in many scenarios, such as passing a signal with white spectrum through an RC lowpass filter.}, which is specified as 
        \begin{equation}
            R_X(t_1, t_2) = R_X(\tau) = P \exp(-\alpha \lvert \tau\rvert),
            \label{eq_theoretical_verification_scheme}
        \end{equation}
        where $\tau = t_1 - t_2$, under AWGN channel with noise PSD being $n_0/2$. The power of signal $X(t)$ is $P=R_X(0)$. According to {\bf Lemma~\ref{lemma_1}}, the trace of the corresponding Mercer operator $M(\cdot)$ is finite. Then the finite-time mutual information given by {\bf Theorem~\ref{th_1}} is also finite, as is shown in {\it Remark~\ref{remark_1}}. Finding the Mercer expansion is equivalent to finding the eigenpairs $(\lambda_k, \phi_k(t))_{k=1}^{\infty}$. The eigenpairs are determined by the following characteristic integral equation \cite{Cai2020Eigenvalue}:
        \begin{equation}
            \lambda_k \phi_k(s) = \int_{0}^{s}{P e^{-\alpha (s-t)} \phi_k(t) \mathrm{d}t}+\int_{s}^{T}{P e^{-\alpha (t-s)} \phi_k(t) \mathrm{d}t}.
            \label{eq_integral_eqn}
        \end{equation}
        Differentiating both sides of \eqref{eq_integral_eqn} twice with respect to $s$ yields the boundary conditions and the differential equation that $\lambda_k, \phi_k$ must satisfy:
        \begin{equation}
            \begin{aligned}
                \lambda_k \phi_{k}''(s) & = (\alpha^2 \lambda_k - 2\alpha P)\phi_k(s), 0<s<T, \\
                \phi_k'(0) & = \alpha \phi_k(0), \\
                \phi_k'(T) & = -\alpha \phi_k(T). \\
            \end{aligned}
        \end{equation}
        Let $\omega_k>0$ denote the resonant frequency of the above harmonic oscillator equation, and let $\phi_k(t)=A_k \cos(\omega_k t) + B_k \sin(\omega_k t)$ be the sinusoidal form of the eigenfunction. Using the boundary conditions we obtain
        \begin{equation}
            \begin{aligned}
                & B_k \omega_k = \alpha A_k, \\
                & B_k \omega_k \cos(\omega_k T)-A_k \omega_k \sin(\omega_k T)  \\
                & =-\alpha \left( A_k \cos(\omega_k T) + B_k \sin(\omega_k T)\right).  \\
            \end{aligned}
            \label{eq_determine_omega_1}
        \end{equation}
        \begin{figure}[!t]
            \centering
            \includegraphics[width=8.5cm]{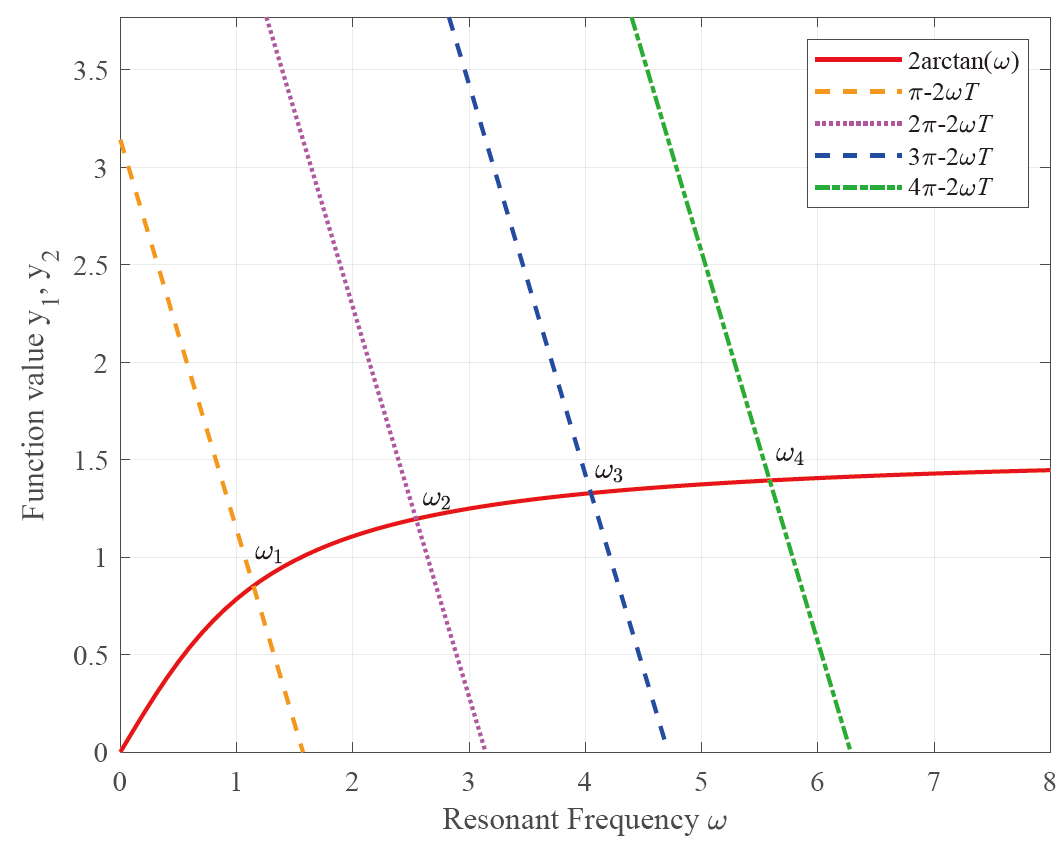}
            \caption{Images of $2\arctan(\omega/\alpha)$ and $k\pi - \omega T$ with $T=2, \alpha=1$ and $k=1,2,3,4$. The desired resonant frequencies $\omega_k$ can be read from the horizontal coordinates of the intersection points of the curve $2\arctan(\omega/\alpha)$ and the parallel lines. }
            \label{fig_eigen_determine_curves}
        \end{figure}
        To ensure the existence of solution to the homogeneous linear equations \eqref{eq_determine_omega_1} with unknowns $A_k, B_k$, the determinant must be zero. Exploiting this condition, we find that $\omega_k$ naturally satisfy the transcendental equation $\tan(\omega_k T) = (2\omega_k \alpha)/({\omega_k^2 -\alpha^2})$. By introducing an auxillary variable $\theta_k = \arctan(\omega_k/\alpha) \in [0, \pi/2]$, this transcendental equation can be simplified as $\tan(\omega_k T) = -\tan(2\theta_k)$, i.e., there exists positive integer $m$ such that $2\arctan(\omega_k / \alpha) = m\pi - \omega_k T$. The integer $m$ can be chosen to be equal to $k$. From the function images of $2\arctan(\omega/\alpha)$ and $k\pi - \omega T$ (Fig.~\ref{fig_eigen_determine_curves}), we can determine $\omega_k$, and then $\lambda_k$. To sum up, the solution to the characteristic equation \eqref{eq_integral_eqn} are collected into \eqref{eq_eigenpairs} as follows.
        \begin{equation}
            \begin{aligned}
                2\arctan(\omega_k /\alpha) & = k\pi - \omega_k T,       \\
                \lambda_k & = \frac{2\alpha P}{\alpha^2 + \omega_k^2},   \\
                \phi_k(t) & = \frac{1}{Z_k} \left(\omega_k \cos(\omega_k t) + \alpha \sin(\omega_k t)\right), \\
            \end{aligned}
            \label{eq_eigenpairs}
        \end{equation}
        where $Z_k$ denotes the normalization constants of $\phi_k(t)$ on $[0,T]$ to ensure orthonormality.

        Equation \eqref{eq_eigenpairs} gives all the eigenpairs $(\lambda_k, \phi_k)_{k=1}^{\infty}$, from which we can calculate $I(T)$ by applying {\bf Theorem~\ref{th_1}}. As for the Shannon limit $C_{\rm sh}$, by applying \eqref{eq_shannon_formula} and evaluating the integral with \cite{integralTable2014} we can obtain
        \begin{equation}
            C_{\rm av} = C_{\rm sh} = \frac{1}{2} \left(\sqrt{\alpha^2+\frac{4P\alpha}{n_0}} - \alpha\right).
            \label{eq_Csh_example2}
        \end{equation}
        After all the preparation works above, we can rigorously prove that $C(T) > C_{\rm av}$ under the typical case of \eqref{eq_theoretical_verification_scheme}, as long as the transmission power $P$ is smaller than a constant $\delta$. The following {\bf Theorem~\ref{th_2}} proves this result.

        \begin{theorem}[Existence of exceed-average phenomenon in a typical case] \label{th_2}
            Suppose $X(t)$ and $Y(t)$ are specified according to \eqref{eq_theoretical_verification_scheme}. The eigenpairs are determined by \eqref{eq_eigenpairs}. Then, for any fixed positive values of $T, n_0$ and $\alpha$, there exists $\delta>0$ such that the exceed-average inequality
            \begin{equation}
                C(T):=\frac{1}{2T}\sum_{k=1}^{+\infty}\log\left(1+\frac{\lambda_k}{n_0/2} \right) > C_{\rm av}
                \label{eq_exceed_shannon_special}
            \end{equation}
            holds strictly for arbitrary $0<P<\delta$.
        \end{theorem}
        \begin{IEEEproof}
            See Appendix A.
        \end{IEEEproof}

        \begin{figure}[!t]
            \centering
            \includegraphics[width=8.5cm]{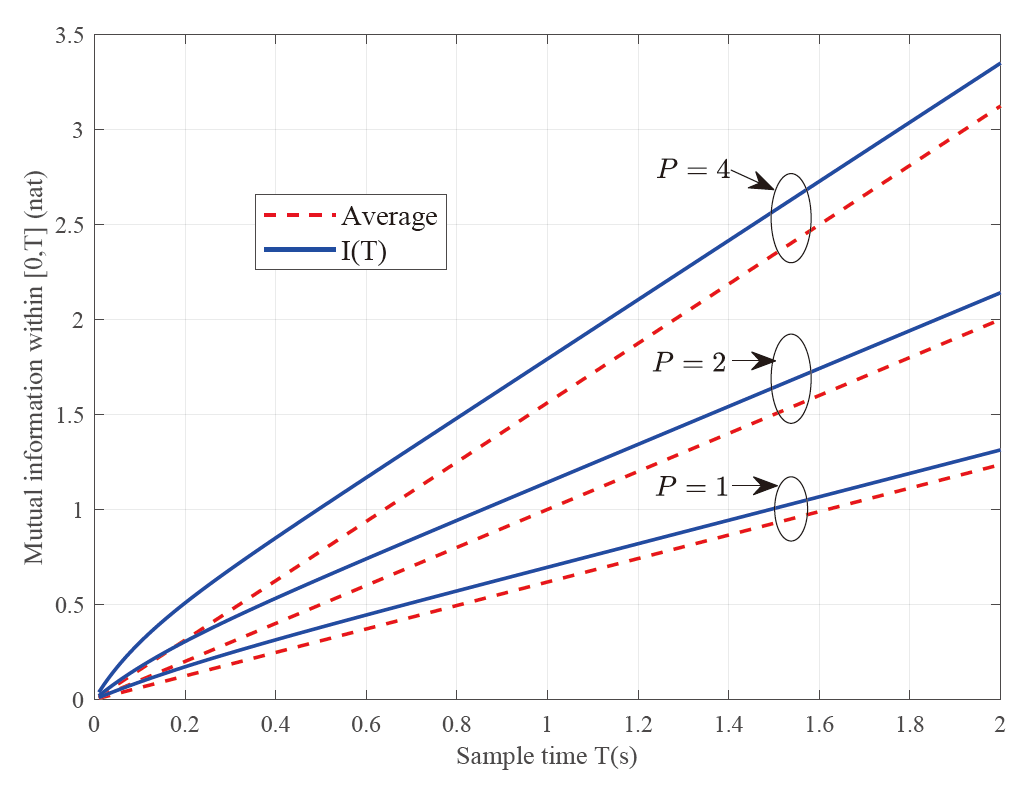}
            \caption{Theoretical verification of the exceed-average effect. The blue lines represent the finite-time mutual information $I(T)$. The red lines are the average mutual information $TC_{\rm av}$, calculated from \eqref{eq_shannon_formula}. All the curves are evaluated under hypothesis \eqref{eq_theoretical_verification_scheme}, where $P=1,2,4$, $n_0 = 1$ and $\alpha=1$.}
            \label{fig_theoretical_exceed_shannon}
        \end{figure}
        To support the above theoretical analysis, numerical experiments on $I(T)$ are conducted based on evaluations of \eqref{eq_eigenpairs} and \eqref{eq_Csh_example2}. As shown in Fig.~\ref{fig_theoretical_exceed_shannon}, it seems that we can always harness more mutual information in a finite-time observation window than the averaged mutual information. This fact is somehow unsurprising because the observations ${\bm Y}(t_1^N)$ inside the window $[0,T]$ can always eliminate some extra uncertainty outside the window due to the autocorrelation of $X(t)$. 

    \subsection{Compatibility with the Shannon-Hartley theorem}\label{sect_4_subsect_2}
        Though the exceed-average effect do imply mathematically that the mutual information within a finite-time window is higher than the average mutual information $C_{\rm av}$ (which coincides with the Shannon limit \eqref{eq_shannon_formula} in this case), in fact, it is still impossible to construct a long-time stable data transmission scheme above the Shannon capacity by leveraging this effect. So the exceed-average phenomenon does not contradict the Shannon-Hartley Theorem. Placing additional observation windows cannot increase the average information rate, because the posterior process $Y(t)|{\bm Y}(t_1^N)$ does not carry as much information as the original one, causing a rate reduction in the later windows. It is expected that, the finite-time mutual information would ultimately decrease to the average mutual information as the total observation time tends to infinity (i.e., $C(\infty)=C_{\rm av}$), and the analytical proof, as well as the achievability of this finite-time mutual information, is still worth investigation in future works.

\section{Conclusions}    
    In this paper, we provided rigorous proofs of the existence of the exceed-average phenomenon under typical autocorrelation settings of the transmitted signal process and the noise process. Our discovery of the exceed-average phenomenon provided a generalization of Shannon's renowned formula $C=W\log(1+S/N)$ to the practical finite-time communications. Since the finite-time mutual information is a more precise estimation of the capacity limit, the optimization target may shift from the average mutual information to the finite-time mutual information in the design of practical communication systems. Thus, it may have guiding significance for the performance improvement of modern communication systems. In future works, general proofs of $C(T)>C_{\rm av}$, independent of the concrete autocorrelation settings, still require further investigation. Moreover, we need to answer the question of how to exploit this exceed-average phenomenon to improve the communication rate, and whether this finite-time mutual information is achievable by a certain coding scheme. In addition, although we have discovered numerically that the finite-time mutual information agrees with the Shannon capacity when $T\rightarrow \infty$, an analytical proof of this result is required in the future.
    % Where does the extra mutual information come from? Is there a systematic method to calculate the finite-time mutual information, without the assumption of the AWGN noise? 

% \section*{Acknowledgment}

\section*{Appendix A \\ Proof Of Theorem \ref{th_2}}
    Plugging \eqref{eq_Csh_example2} into the right-hand side of \eqref{eq_exceed_shannon_special}, and differentiate both sides w.r.t $P$. Notice that if $P=0$, then both sides of \eqref{eq_exceed_shannon_special} are equal to 0. Thus, we only need to prove that the derivative of left-hand side is strictly larger than that of right-hand side within a small interval $P\in (0,\delta)$:
    \begin{equation}
        \frac{1}{2}\sum_{k=1}^{+\infty}{\left(\frac{1}{1+\frac{2\lambda_k}{n_0}}\frac{2\lambda_k}{n_0P}\right)} > \frac{T}{n_0}\frac{1}{\sqrt{1+4P/(n_0\alpha)}}.
        \label{eq_exceed_shannon_special_1}
    \end{equation}
    Multiply both sides of \eqref{eq_exceed_shannon_special_1} by $n_0$ and define $\mu_k:=\lambda_k/(PT)$, and then from {\bf Lemma \ref{lemma_1}} we obtain $\sum_k{\mu_k}=1$. In this way,  \eqref{eq_exceed_shannon_special_1} is equivalent to
    \begin{equation}
        \sum_{k=1}^{+\infty}{\frac{\mu_k}{1+\frac{2\lambda_k}{n_0}}} > \frac{1}{\sqrt{1+4P/(n_0\alpha)}}.
        \label{eq_exceed_shannon_special_2}
    \end{equation}
    Since $\varphi(x):=1/(1+2x/n_0)$ is convex on $(0,+\infty)$, by applying Jensen's inequality to the left-hand side  of \eqref{eq_exceed_shannon_special_2}, we only need to prove that
    \begin{equation}
        \frac{1}{1+\frac{2}{n_0}\sum_k{\lambda_k\mu_k}} > \frac{1}{\sqrt{1+4P/(n_0\alpha)}}.
        \label{eq_after_jensen}
    \end{equation}
    From the definition of $\mu_k$ we can derive that  $\lambda_k \mu_k = \lambda_k^2/(PT)$. So we go on to calculate $\sum_k{\lambda_k^2}$. That is equivalent to calculate $\mathrm{tr}(M^2)$, where $M^2$ corresponds to the integral kernel:
    \begin{equation}
        K_{M^2}(t_1, t_2) := \int_{0}^{T}{P^2\exp(-\alpha\lvert t_1-s\rvert)\exp(-\alpha\lvert t_2-s\rvert)\mathrm{d}s}.
    \end{equation}
    Evaluating the kernel $K_{M^2}$ on the diagonal $t=t_1=t_2$, and integrating this kernel on the diagonal of $[0,T]^2$ gives $\sum_{k}{\lambda_k^2}$, i.e., $\mathrm{tr}(M^2)$:
    \begin{equation}
        \begin{aligned}
            \sum_k{\lambda_k^2} & = \frac{P^2}{2\alpha}\int_{0}^{T}{\left(2-e^{-2\alpha t}-e^{-2\alpha (T-t)} \right) \mathrm{d}t}, \\
            & = \frac{P^2}{\alpha}\left(T - \frac{1}{2\alpha}(1-e^{-2\alpha T})\right). \\
        \end{aligned}
        \label{eq_exceed_shannon_special_3}
    \end{equation}
    By substituting \eqref{eq_exceed_shannon_special_3} into \eqref{eq_after_jensen}, we just need to prove that
    \begin{equation}
        \sqrt{1+4P/(n_0\alpha)} > 1+\frac{2P}{n_0\alpha}\left(1-\frac{1-e^{-2\alpha T}}{2\alpha T}\right).
        \label{eq_exceed_shannon_special_4}
    \end{equation}
    Define the dimensionless number $x=2P/(n_0\alpha)$. Since the function $\psi(x):=(1-\exp(-x))/x$ is strictly positive and less than 1 at $x>0$, we can conclude that, there exists a small positive $\delta>0$ such that \eqref{eq_exceed_shannon_special_4} holds for $0<P<\delta$. The number $\delta$ can be chosen as
    \begin{equation}
        \delta = \frac{n_0\alpha\psi(2\alpha T)}{(1-\psi(2\alpha T))^2} > 0,
    \end{equation}
    which implies that \eqref{eq_exceed_shannon_special_1} holds for any $0<P<\delta$. Thus, integrating \eqref{eq_exceed_shannon_special_1} on both sides from $p=0$ to $p=P, P<\delta$ gives rise to the conclusion \eqref{eq_exceed_shannon_special}, which completes the proof of {\bf Theorem \ref{th_2}}.

\section*{Acknowledgment}
This work was supported in part by the National Key Research and Development Program of China (Grant No.2020YFB1807201), in part by the National Natural Science Foundation of China (Grant No. 62031019), and in part by the European Commission through the H2020-MSCA-ITN META WIRELESS Research Project under Grant 956256.

\clearpage
\IEEEtriggeratref{6}
\bibliography{IEEEabrv, refs}

\begin{thebibliography}{10}
\providecommand{\url}[1]{#1}
\csname url@samestyle\endcsname
\providecommand{\newblock}{\relax}
\providecommand{\bibinfo}[2]{#2}
\providecommand{\BIBentrySTDinterwordspacing}{\spaceskip=0pt\relax}
\providecommand{\BIBentryALTinterwordstretchfactor}{4}
\providecommand{\BIBentryALTinterwordspacing}{\spaceskip=\fontdimen2\font plus
\BIBentryALTinterwordstretchfactor\fontdimen3\font minus
  \fontdimen4\font\relax}
\providecommand{\BIBforeignlanguage}[2]{{%
\expandafter\ifx\csname l@#1\endcsname\relax
\typeout{** WARNING: IEEEtran.bst: No hyphenation pattern has been}%
\typeout{** loaded for the language `#1'. Using the pattern for}%
\typeout{** the default language instead.}%
\else
\language=\csname l@#1\endcsname
\fi
#2}}
\providecommand{\BIBdecl}{\relax}
\BIBdecl

\bibitem{Shannon}
C.~E. Shannon, ``A mathematical theory of communication,'' \emph{The Bell Syst.
  Techni. J.}, vol.~27, no.~3, pp. 379--423, Jul. 1948.

\bibitem{Sampling1967Landau}
H.~Landau, ``Sampling, data transmission, and the nyquist rate,'' \emph{Proc.
  {IEEE}}, vol.~55, no.~10, pp. 1701--1706, Oct. 1967.

\bibitem{Nyquist1928CertainTopics}
H.~Nyquist, ``Certain topics in telegraph transmission theory,'' \emph{Trans.
  American Institute of Electrical Engineers}, vol.~47, no.~2, pp. 617--644,
  Apr. 1928.

\bibitem{BW1976Slepian}
D.~Slepian, ``On bandwidth,'' \emph{Proc. {IEEE}}, vol.~64, no.~3, pp.
  292--300, Mar. 1976.

\bibitem{Cover1999ElementsInfTheory}
T.~M. Cover, \emph{Elements of information theory}.\hskip 1em plus 0.5em minus
  0.4em\relax John Wiley \& Sons, 1999.

\bibitem{mercer1909functions}
J.~Mercer, ``Functions of positive and negative type and their connection with
  the theory of integral equations,'' \emph{Philos. Trans. Royal Soc.}, vol.
  209, pp. 4--415, 1909.

\bibitem{zhu2007operator}
K.~Zhu, \emph{Operator theory in function spaces}.\hskip 1em plus 0.5em minus
  0.4em\relax American Mathematical Soc., 2007, no. 138.

\bibitem{GPSampling1957BA}
A.~Balakrishnan, ``A note on the sampling principle for continuous signals,''
  \emph{IRE Trans. Inf. Theory}, vol.~3, no.~2, pp. 143--146, Jun. 1957.

\bibitem{MercerExpansion1955Barrett}
J.~Barrett and D.~Lampard, ``An expansion for some second-order probability
  distributions and its application to noise problems,'' \emph{IRE Trans. Inf.
  Theory}, vol.~1, no.~1, pp. 10--15, Mar. 1955.

\bibitem{Cai2020Eigenvalue}
D.~Cai and P.~S. Vassilevski, ``Eigenvalue problems for exponential-type
  kernels,'' \emph{Comput. Methods in Applied Math.}, vol.~20, no.~1, pp.
  61--78, Jan. 2020.

\bibitem{integralTable2014}
I.~S. Gradshteyn and I.~M. Ryzhik, \emph{Table of integrals, series, and
  products}.\hskip 1em plus 0.5em minus 0.4em\relax Academic press, 2014.

\end{thebibliography}

\end{document}